\documentclass[aip,sd,amsmath,amssymb,reprint]{revtex4-1}

\usepackage{graphicx}
\usepackage{dcolumn}
\usepackage{bm}

\begin{document}

\newcommand{\eps}{\varepsilon}

\title{Interacting nonlinear wave envelopes and rogue wave formation in deep water}

\author{Mark J. Ablowitz}
\affiliation{Department of Applied Mathematics, University of Colorado, 526
UCB, Boulder, CO 80309-0526, USA}

\author{Theodoros P. Horikis}\email{Corresponding author: horikis@uoi.gr}
\affiliation{Department of Mathematics, University of Ioannina, Ioannina
45110, Greece}

\date{\today}

\begin{abstract}
A rogue wave formation mechanism is proposed within the framework
of a coupled nonlinear Schr\"{o}dinger (CNLS) system corresponding
to the interaction of two waves propagating in oblique directions
in deep water. A {\it rogue condition} is introduced that links the
angle of interaction with the group velocities of these waves:
different angles of interaction can result in a major enhancement
of rogue events in both numbers and amplitude. For a range of
interacting directions it is found that the CNLS system exhibits
significantly more extreme wave amplitude events than its scalar
counterpart. Furthermore, the rogue events
of the coupled system are found to be well approximated
by hyperbolic secant functions; they are vectorial soliton-type solutions of the CNLS system, typically not considered to be integrable.
Overall, our results indicate that crossing states provide an
important mechanism for the generation of rogue water wave events.

\end{abstract}


\maketitle

Even in the age of modern ship building and innovative navigation
equipment, ship accidents are an important area of maritime
concern. In recent years researchers have been investigating a
topic which heretofore had been reserved for marine folklore: giant
waves that seem to appear from nowhere in high seas that can lead
to disastrous outcomes. These waves have, nowadays, been documented
to exist and are usually referred to as rogue or freak waves.

The conditions that cause such waves to grow enormously in size are a topic
of great interest with many different hypotheses being proposed
\cite{book,waters}. An important one is the nonlinear mechanism of the
self-wave interactions, such as modulation instability; in water wave physics
this is called the Benjamin-Feir instability \cite{bf} and has been found
under certain conditions to produce significant wave amplification.

The envelope of nonlinear water waves, under suitable conditions, is modeled
by a nonlinear Schr\"{o}dinger equation (NLS) \cite{waves}. Interestingly
enough, the NLS not only gives a suitable description of these waves
\cite{dysthe,muller,waters,chabchoub,grimshaw,shats}, it is also an important
equation used to investigate propagation of pulses in many other physical
systems such as nonlinear optical fibers
\cite{kodama,pisarchik,bonatto,zaviyalov}, Bose-Einstein condensates
\cite{bec1}, magnetic spin waves \cite{patton} and many others.

The state of the sea in which rogue waves form is often complex
\cite{onorato2,onorato3,ruban3} with certain key wave interactions dominating
the wave structure. Here we focus on a coupled NLS (CNLS) system, derivable
directly from the Euler equations. Underlying the wave phenomena are separate
wave trains, propagating in different directions which interact at different
angles; this provides a useful model of crossing sea states
\cite{adcock,toffoli,cavaleri,pearson,onorato2}. The scalar NLS equation
would be insufficient to describe these phenomena (rogue events associated
with 2D scalar NLS equations are discussed in Ref. \cite{buvoli}).

As with the scalar problem the CNLS system exhibits modulation instability
(MI), which as mentioned above, is an important aspect of rogue wave
dynamics. Indeed, recent optical experiments indicate that MI is  critical in
the generation of strong optical rogue wave phenomena \cite{solli}. Previous
studies have considered the growth rate of the underlying MI associated with
CNLS equations \cite{onorato2}. It was found \cite{kourakis,onorato2} that
the CNLS system can have significantly higher growth rates than the scalar
NLS equation and thus MI will be triggered sooner. Here we find that for
certain angles of wave interaction the CLNS system leads to large amplitude
rogue waves and there can be more events than in the scalar NLS analogue. For
other angles of interaction, we find another interesting result: while the
growth rate can be  higher in the CNLS system, it does not
necessarily result in more
significant events than its scalar counterpart.

Modulational instability and the wave angles of interaction of the waves in
the CNLS system are directly linked to the coefficients of the system. The
strongest rogue wave effect is found when both equations in the CNLS system
are of focusing type. In fact, for special values of these coefficients the
CNLS is integrable \cite{manakov}. For the scalar NLS equation, when
dispersion and nonlinearity share the same signs, the equation is said to be
focusing, and the equation is modulationaly unstable; it is stable when the
signs are opposite and the equation is said to be defocusing. The stability
criteria are more involved for the coupled system \cite{kourakis}.

Here, we introduce the concept of a {\it rogue condition} which allows the
two (spatial) dimensional water wave amplitude system to be reduced to a
family of one dimensional CNLS systems. Indeed rogue events of associated
one-dimensional equations such as the NLS equation has been a major topic of
research  in the study of rogue waves.  It is found that there is a direct
link between the rogue condition, i.e. the angle of interaction, and the number
and magnitude of rogue events. We show that  for certain angles this has a major
effect on the associated wave events. Thus this rogue condition is not only related to
the
MI mechanism but also to the number and severity of events. A key aspect of this
work is to investigate this rogue condition via direct calculation of
probability distributions. We do not discuss the details of the
underlying  statistics which a separate topic and is outside the scope
of this paper. For our purposes a basic probability density function
distribution explains the salient points.

The nature of these rogue events is also an important issue addressed here. Rogue
waves are commonly associated in the literature with rational-type solutions \cite{peregrine,breather} of the scalar integrable NLS equation. We show that the rogue events associated with the CNLS equation are
well approximated by hyperbolic secant functions. They propagate as  solitary wave or soliton-type solutions
of the CNLS system, which is not known to be integrable. Integrable equations exhibit  rational solutions which are limits of multi-soliton solutions \cite{ist} which by their very nature are associated with integrable systems like the scalar NLS equation.

Since we will be analyzing coupled NLS equations that arise in water waves it
is convenient to review the analysis beginning with the Euler equations
appropriate for describing waves in deep water
\begin{gather*}
\nabla^2\phi=0, \quad-\infty < z < \eps\eta\\
\phi_z=0,\quad z\rightarrow-\infty \\
\eta_t+\eps(\eta_x\phi_x+\eta_y\phi_y)=\phi_z,\quad z = \eps\eta\\
\phi_t+g\eta+\frac{\eps}{2}(\phi_x^2+\phi_y^2+\phi_z^2)=0, \quad z=\eps\eta
\end{gather*}
where $\phi$ is the velocity potential, $\eta$ the surface elevation, $g$
gravity, $\boldsymbol{\nabla}=(\partial/\partial x,\partial/\partial y)$ and
$\eps$ is a small parameter. Define the multiple scales $ X=\eps x$, $Y=\eps
y$, $Z=\eps z$, $T= \eps t$, expand $\phi=\phi(t,x,y,\eps\eta)$ around $z=0$,
and all fields in powers of $\eps$ as follows
\begin{gather*}
\phi=\phi_0+\eps\phi_1+\mathcal{O}(\eps^2)\\
\eta=\eta_0+\eps\eta_1+\mathcal{O}(\eps^2).
\end{gather*}
At leading order, $\mathcal{O}(1)$, we assume two wave trains
\begin{subequations}
\begin{align}
\phi_0 &=A(X,Y,Z,T)e^{i\theta_1+\Omega_1 z}
+B(X,Y,Z,T)e^{i\theta_2+\Omega_2 z}+\mathrm{cc}\\
\eta_0 &=N(X,Y,T)e^{i\theta_1}+M(X,Y,T)e^{i\theta_2}+\mathrm{cc}\label{eta}
\end{align}
\end{subequations}
where $\theta_i=k_i x+l_i y-\omega_i t$, $(i=1,2)$,
$\Omega_i=\sqrt{k_i^2+l_i^2}$, $\omega_i^2=g\sqrt{k_i^2+l_i^2}$ and cc
denotes complex conjugate. Using the above and removing secular terms the
following set of coupled equations is obtained
\begin{subequations}
\begin{gather}
i \left( {\frac{{\partial {A}}}{{\partial {T}}} + {{\mathbf{v}}_1} \cdot
\boldsymbol{\nabla}
{A}}
\right) +
\eps \left( \frac{1}{2}\frac{{{\partial
^2}{\omega _1}}}{{\partial k_1^2}}\frac{{{\partial ^2}{A}}}{{\partial X^2}} +
\frac{{{\partial ^2}{\omega
_1}}}{{\partial
{k_1}\partial {l_1}}}\frac{{{\partial ^2}{A}}}{{\partial {X}\partial {Y}}}\right.
\nonumber\\
+ \left.\frac{1}{2}\frac{{{\partial ^2}{\omega _1}}}{{\partial l_1^2}}\frac{{{\partial
^2}{A}}}{{\partial Y^2}}
+
({q_{11}}|{A}{|^2} + {q_{12}}|{B}{|^2}){A} \right) = 0  \hfill \\
i \left( {\frac{{\partial {B}}}{{\partial {T}}} + {{\mathbf{v}}_2} \cdot
\boldsymbol{\nabla}
{B}}
\right) +
\eps \left(\frac{1}{2}\frac{{{\partial
^2}{\omega _2}}}{{\partial k_2^2}}\frac{{{\partial ^2}{B}}}{{\partial X^2}} +
\frac{{{\partial ^2}{\omega
_2}}}{{\partial
{k_2}\partial {l_2}}}\frac{{{\partial ^2}{B}}}{{\partial {X}\partial {Y}}}\right.
\nonumber\\
+\left. \frac{1}{2}\frac{{{\partial ^2}{\omega _2}}}{{\partial l_2^2}}\frac{{{\partial
^2}{B}}}{{\partial Y^2}}
+ ({q_{21}}|{A}{|^2} + {q_{22}}|{B}{|^2}){B} \right)= 0 \hfill
\end{gather}
\label{cnls1}
\end{subequations}
where $\mathbf{v}_i=(\partial\omega_i/\partial k_i,\partial\omega_i/\partial
k_i)^\mathrm{T}$ and
\begin{gather*}
q_{11}=-\frac{2\omega_1^7}{g^4},\quad q_{22}=-\frac{2\omega_2^7}{g^4}\\
q_{12}=-\frac{1}{\omega_1}\left[ m_{12}^2 +\frac{4m_{12}}{g^2}\omega_1\omega_2^3
-\frac{\omega_1^4\omega_2^4}{g^4} \right]\\
q_{21}=-\frac{1}{\omega_2}\left[ m_{12}^2 +\frac{4m_{12}}{g^2}\omega_1^3\omega_2
-\frac{\omega_1^4\omega_2^4}{g^4} \right]
\end{gather*}
where $m_{12}= k_1k_2+l_1l_2$. These coefficients are in agreement with the
ones found in Refs. \cite{benney,roskes,segur}. Indeed, the  nonlinear
coefficients all agree because they do not depend on derivatives of the
fields $A$, $B$. Similarly, the coefficients of the linear terms are
associated with the underlying linear dispersion relation of water waves;
this is universal for NLS systems.

The relationship between the velocity potential $\phi$ and the wave elevation
$\eta$ is obtained via
\[
N=\frac{i\omega_1}{g}A,\quad M=\frac{i\omega_2}{g}B.
\]

Next introduce new {\it projection} coordinates such that
\begin{gather*}
\xi_1=\cos\theta X+\sin\theta Y-
\left( \cos\theta \frac{\partial \omega_1}{\partial k_1} +\sin\theta
\frac{\partial\omega_1}{\partial l_1}
\right)T\\
\xi_2=\cos\theta X+\sin\theta Y-
\left( \cos\theta \frac{\partial \omega_2}{\partial k_2} +\sin\theta
\frac{\partial\omega_2}{\partial l_2}
\right)T
\end{gather*}
and demand that we have only one projected coordinate, $\xi_1=\xi_2=\xi$,
i.e.
\begin{gather}
\cos\theta \frac{\partial \omega_1}{\partial k_1} +\sin\theta
\frac{\partial\omega_1}{\partial l_1} =
\cos\theta \frac{\partial \omega_2}{\partial k_2} +\sin\theta
\frac{\partial\omega_2}{\partial l_2}
\Leftrightarrow\nonumber \\
\tan\theta=-\frac{\partial \omega_1/\partial k_1-\partial \omega_2/\partial
k_2}{\partial
\omega_1/\partial
l_1-\partial
\omega_2/\partial l_2}
\label{rogue}
\end{gather}
We refer to Eq. \eqref{rogue}, which links the angle of interaction with the
wave numbers of the relative waves, as the {\it rogue condition}. Note, that
if more than two wave packets are present this condition is much more
restrictive as the resulting system of equations will be satisfied only for
certain wave numbers \cite{roskes2}. It is important to observe that
transforming to projection coordinates, with the above rogue condition added,
has the effect of matching the group velocities of the different cross wave
states. If the group velocities were not matched then two localized states
would evolve through one another and lead to minimal interaction. The rogue
condition thus allows enhanced interaction.

This condition leads to a family of one dimensional coupled NLS systems all
of which have the potential to generate rogue waves. This is more general
than the case studied in Ref. \cite{proment} where wave numbers were chosen
such that $k_1=k_2$ and $l_1=-l_2$. As such only one case whose projection
moves along one preferred axis was considered. Furthermore, only the growth
rates of modulation instability were considered and no comparisons between
the NLS and the number and frequency of coupled systems's rogue event
statistics were made. Demonstrating statistically that more frequent and more
serious rogue events occur in the coupled system shows that surely these
systems must be considered as potential rogue event generators. We also note
that our envelope equations are different from the those of Ref.
\cite{onorato2} which are derived from an approximation to the Euler
equations. The equations here are derived directly from the Euler
equations.

Moreover, the nonlinear Schr\"odinger equation (in 1+1: one space, one time
dimension or even 2+1
dimensions) corresponds to wave packets --envelopes associated with a
periodic wave train which is slowly varying in space.  In the 1+1 NLS
equation one considers only a one dimensional model along the direction of a
wave train. In 2+1 one requires slow variation in both the direction of wave
train and perpendicular to the wave train; in essence since the envelope is
slowly varying, the 2+1 model itself can be considered as essentially
quasi-one dimensional.

In general one can ask whether a sea state can be one dimensional. Indeed
nearly one dimensional sea states are special situation but nearly one
dimensional states can occur, and while these are somewhat unusual or special
situations nevertheless the issue under study here is, by its very nature,
special: i.e. a rogue event. One can also consider slowly varying waves both
along and orthogonal to the direction to the wave train \cite{buvoli}. As
indicated above, without the one dimensional or nearly one dimensional
reduction one expects that any localized disturbance in coupled NLS equations
would pass through one another and there would be a relatively small
resulting interaction.

Finally, with $\tau=\eps T$  Eqs. \eqref{cnls1} become
\begin{subequations}
\begin{gather}
  i\frac{{\partial {A}}}{{\partial \tau }} + {P_1}\frac{{{\partial ^2}{A}}}{{\partial
  {\xi
  ^2}}} +
  ({q_{11}}|{A}{|^2} +
  {q_{12}}|{B}{|^2}){A} = 0 \hfill \\
  i\frac{{\partial {B}}}{{\partial \tau}} + {P_2}\frac{{{\partial ^2}{B}}}{{\partial
  {\xi
  ^2}}} +
  ({q_{21}}|{A}{|^2} +
  {q_{22}}|{B}{|^2}){B} = 0
  \end{gather}
 \label{roguemodified}
\end{subequations}
where $\displaystyle P_i=\frac{1}{2}\left[ {{{\cos }^2}\theta
\frac{{{\partial ^2}{\omega _i}}}{{\partial k_i^2}} + \sin (2\theta)
\frac{{{\partial ^2}{\omega _i}}}{{\partial {k_i}\partial {l_i}}} + {{\sin
}^2}\theta \frac{{{\partial ^2}{\omega _i}}}{{\partial l_i^2}}}\right]$.
Notably, Eqs. \eqref{roguemodified} have the same structure in dimensionless
form. Indeed, introduce the nondimensional scaling:  $k_j= k_* k'_j$, $l_j=
k_* l'_j$, $j=1,2$ and $\tau' = \omega_* \tau$, $\xi'= k_* \xi$, $A' =N_*A$,
$B' =N_*B$, where $N_*= g/(\omega_* / k_*)$, $\omega_*^2= g k_*$ where $k_*$
is a typical wavelength in the ocean. The result, after dropping primes, is
exactly the system of equations Eqs. \eqref{roguemodified} above.

\begin{figure}[!htbp]
    \centering
    \includegraphics[width=2.5in]{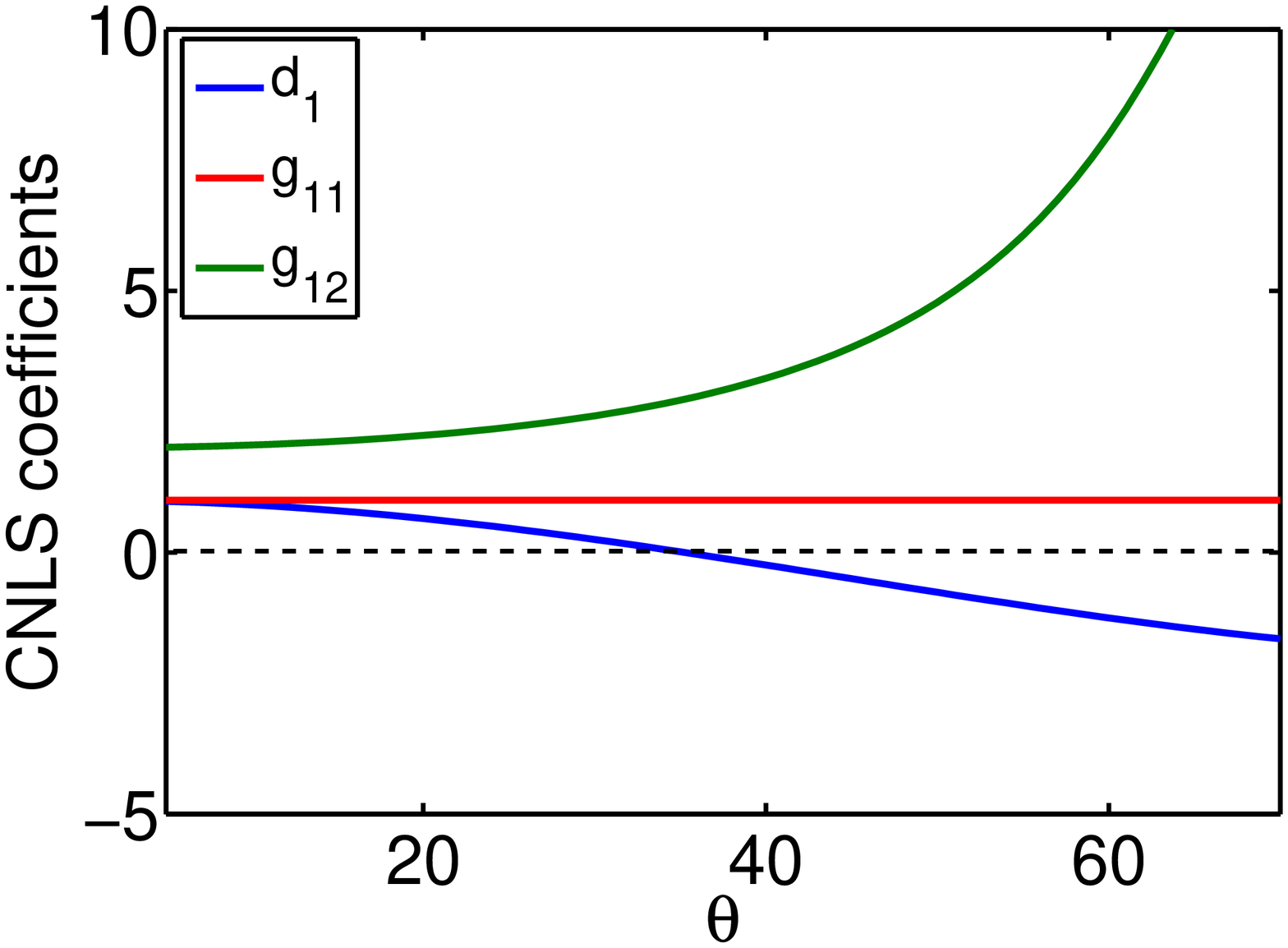}
    \includegraphics[width=2.5in]{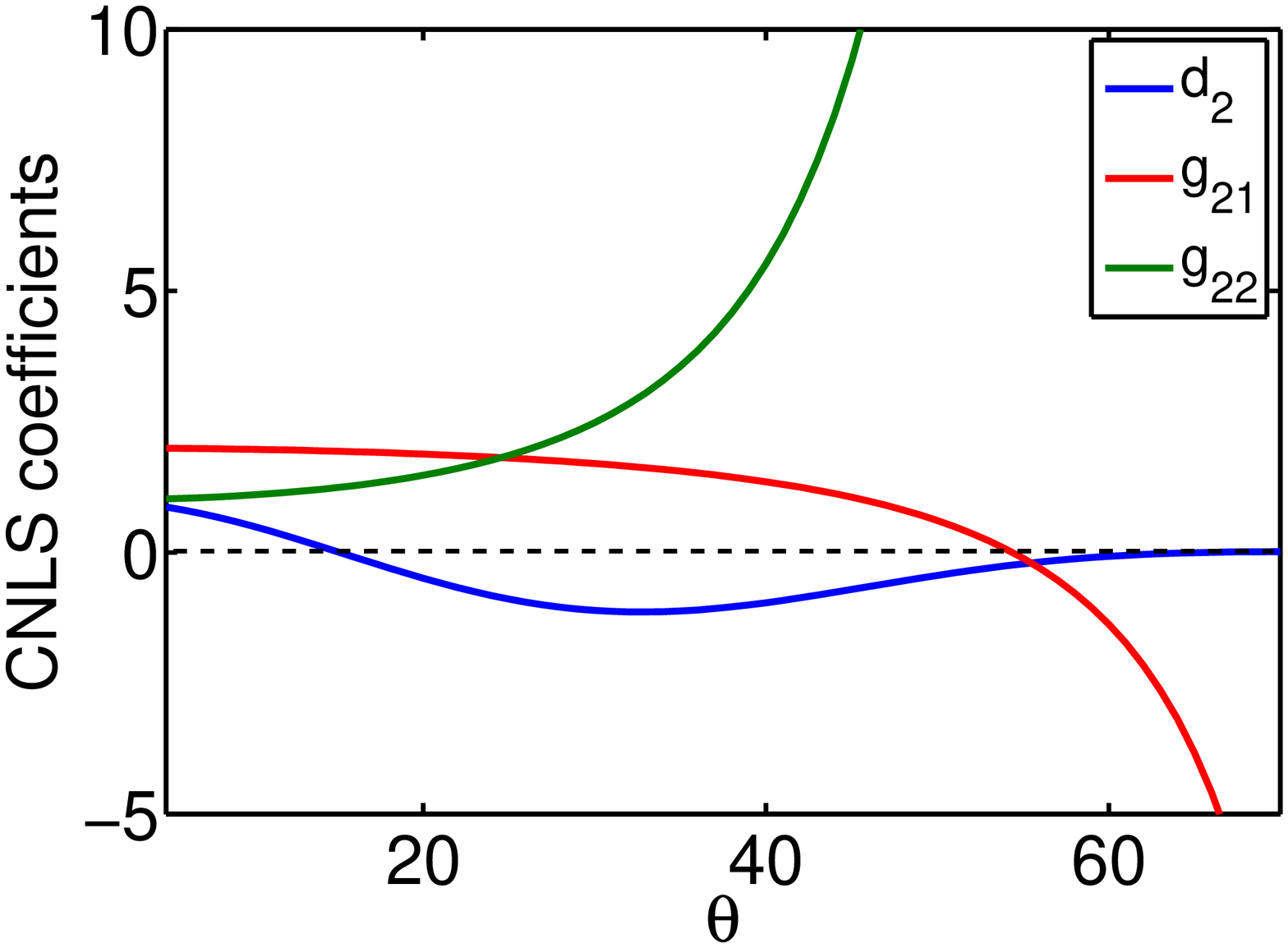}
    \caption{(Color online) The CNLS coefficients vs. the angle $\theta$.}
    \label{coeff}
\end{figure}

As a prototypical situation in what follows we fix $k_1=k_2=1$ and $l_1=0$
and vary the angle $\theta$ so that $l_2$ is retrieved from Eq. \eqref{rogue}.
Furthermore, for convenience, introduce the change of variables:
\[
\tau= \tau_0\tau',\; \xi= \xi_0\xi',\; A= A_0 A',\;
B= B_0 B'
\]
where, $\tau_0=-2 \omega_1$, $\xi_0= \omega_1/(2 k_1)$ and $A_0=B_0=
1/(2k_1^2)$. Then after dropping primes, we obtain the  system
\begin{subequations}
\begin{align}
i A_{\tau}+d_1 A_{\xi\xi}+(g_{11}|A|^2+g_{12}|B|^2) A &= 0
\label{cnls.final1}\\
i  B_{\tau}+d_2 B_{\xi\xi}+\left(g_{21}|A|^2+g_{22}|B|^2\right) B &= 0
\label{cnls.final2}
\end{align}
\label{cnls.final}
\end{subequations}
With the above values the coefficients of the system \eqref{cnls.final} vary
with $\theta$ as shown in Fig. \ref{coeff}.

We note that at about $\theta=35^\circ$ the zero dispersion limit of Eq.
\eqref{cnls.final1} occurs ($d_1=0$) and similarly for Eq.
\eqref{cnls.final2} at about $\theta=15^\circ$ and $65^\circ$ ($d_2=0$); here
the above rescaling fails and a different scaling must be used. Finally note
that as $\theta\rightarrow 90^\circ$ $|g_{21}| \rightarrow \infty$ and the
equations uncouple with $A\rightarrow 0$.

\begin{figure}[!htbp]
    \centering
    \includegraphics[width=2.5in]{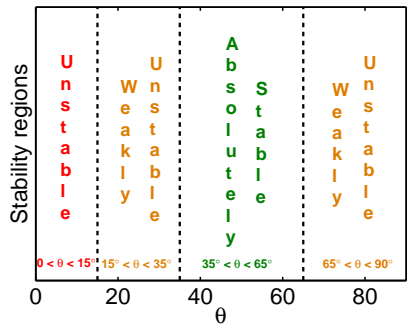}
        \caption{(Color online) The three stability regions of Eqs.
        \eqref{cnls.final}.}
    \label{stability}
\end{figure}

The sign changes of the coefficients indicated in Fig. \ref{coeff} also lead
to changes in the modulational stability properties for Eqs.
\eqref{cnls.final}; this is summarized in Fig. \ref{stability}. It is
important to stress that growth rates and  occurrence of rogue events and are
not necessarily indicative of the wave statistics. Indeed, based on the results
of Ref. \cite{kourakis} the coupled system exhibits MI  much  faster than its
scalar counterpart, i.e. has higher growth rates. This, however, may not
result in more extreme events as will be shown  here.

We name these regions unstable (where both equations are focusing), weakly
unstable (where one of the equations is focusing and the other defocusing)
and absolutely stable (where both equations are defocusing). Recall
\cite{kourakis}, that the stability criteria are different for the scalar and
coupled NLS systems. However, if both equations are defocusing the CNLS is
always stable.

We now integrate numerically Eqs. \eqref{cnls.final} using pseudospectral
methods in space and  exponential Runge-Kutta for the evolution Ref.
\cite{kassam} in a computational domain $\xi \in[-100,100]$, $\tau\in[0,10]$.
As a convenient and ``typical" initial condition we choose a wide gaussian
\[
A(\xi,0)= B(\xi,0)=e^{-\xi^2/2 \sigma^2},\; \sigma=30
\]
perturbed with 10\% random noise for $10^5$ trials. A wide gaussian with
randomness added is a prototype of a  set of broad/randomly generated states.
The Gaussian initial data we chose also has narrow band spectrum since we are
looking at slowly varying waves around two cross wave. We can modify the
distribution to look more like JONSWAP spectrum data (see below), but  the
essential purpose here, as indicated above  is to show that the coupled
system can lead to serious rogue events similar to, but more serious than the
scalar NLS equation.

\begin{figure}[tbp]
    \centering
    \includegraphics[width=2.5in]{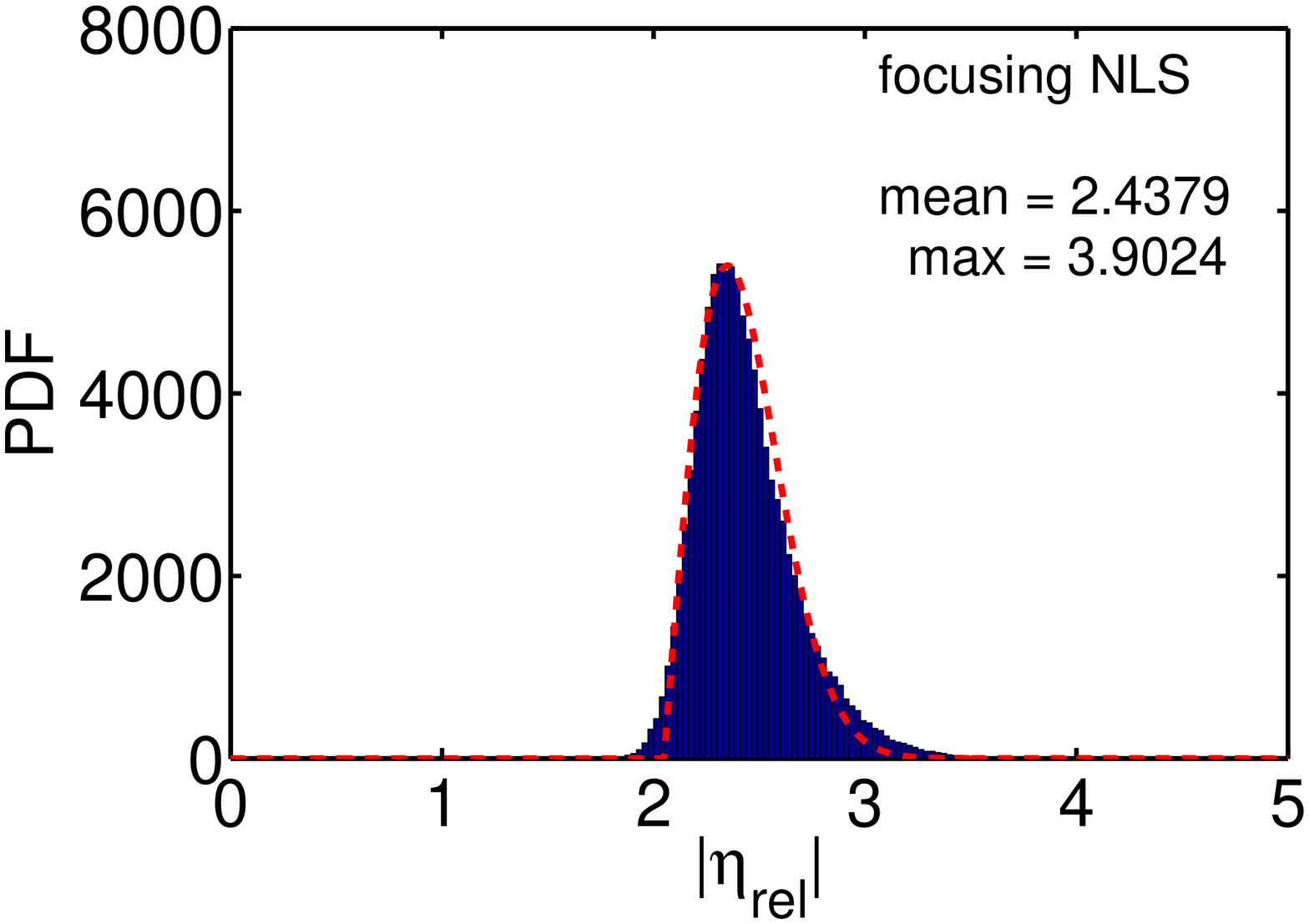}
    \includegraphics[width=2.5in]{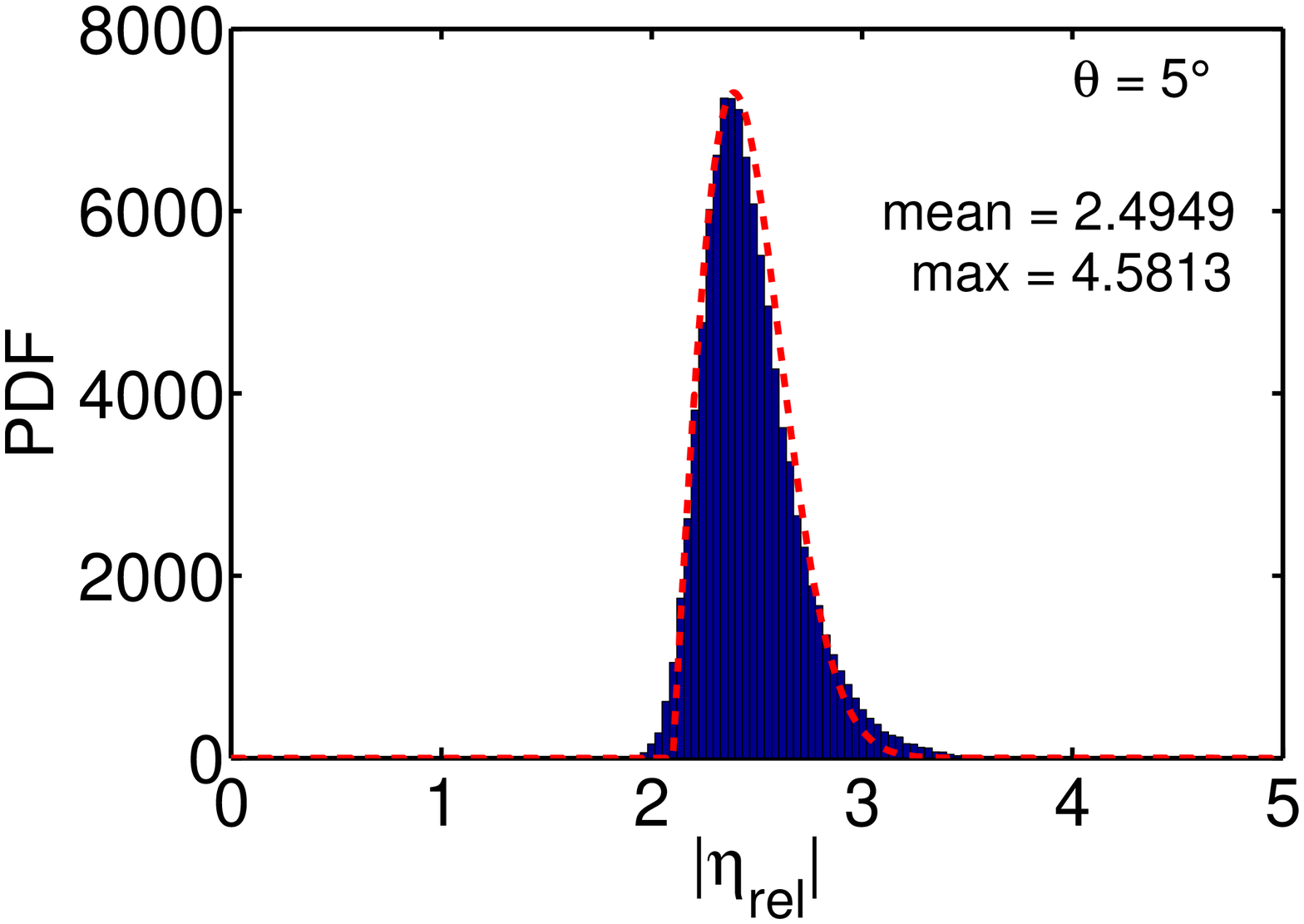}
    \includegraphics[width=2.5in]{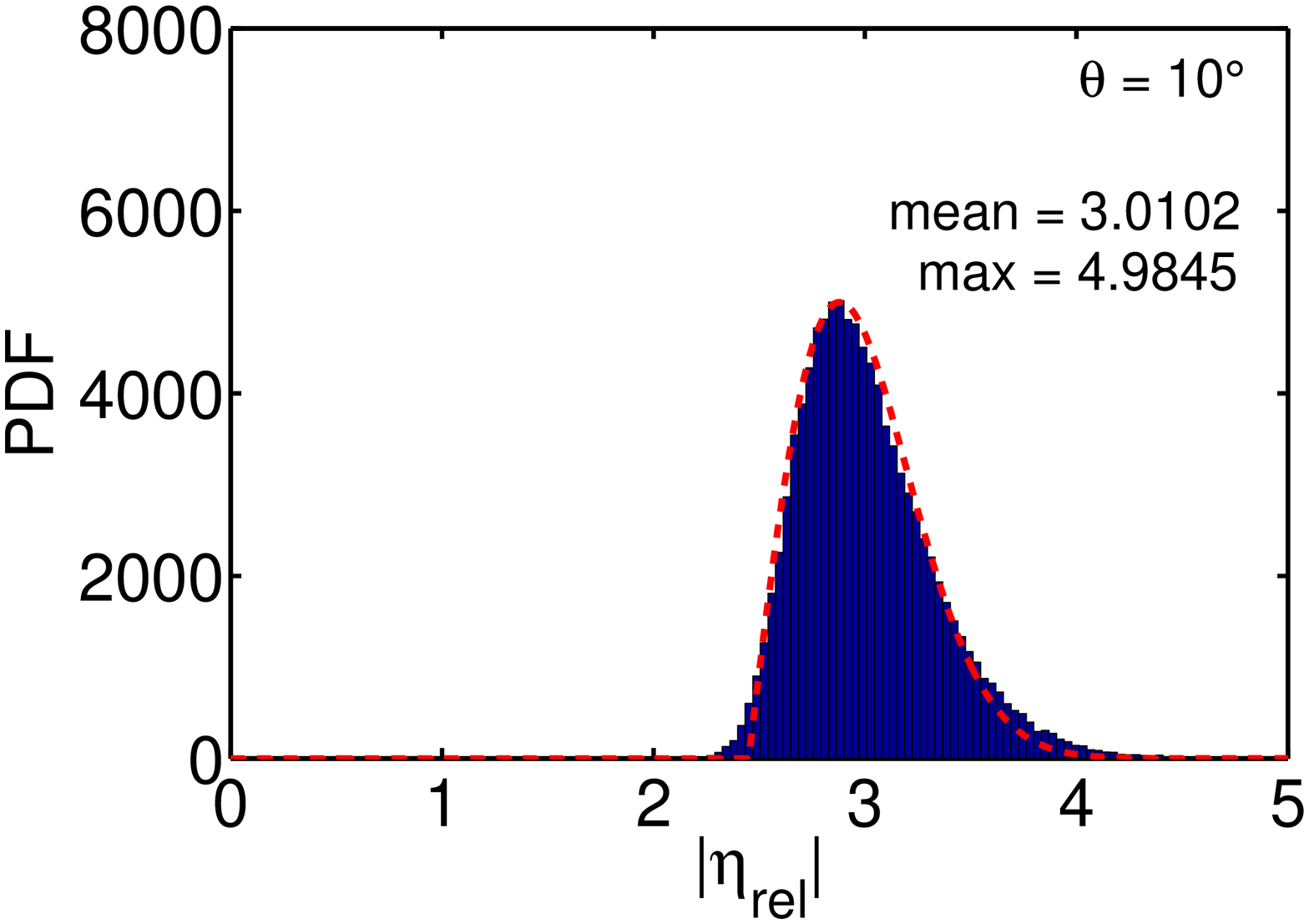}
    \caption{(Color online) The PDFs for the scalar focusing NLS (top) and unstable
    CNLS.
    A Rayleigh distribution with the same peak and scale factor as our
    calculated PDF is fitted (in red) on each PDF.}
    \label{pdf1}
\end{figure}

In each trial we measure the highest wave amplitude and introduce the
quantity $\eta_{\mathrm{rel}}$ which is the ratio of the highest wave
amplitude (as defined by Eq. \eqref{eta}) to the maximum of the initial
condition. In the following figures we show the resulting probability density
functions (PDF) for the two different regions of instability (in the stable
region we see rogue events in neither the CLNS system nor the scalar
defocusing NLS equation). It is found that the PDFs exhibit Rayleigh-like
distributions (see Fig. \ref{pdf1}) in line with our evaluation of the
magnitude of the wave elevation. We do not go into the details of the
differences between our distribution and Rayleigh distributions as such
differences are not an essential aspect of this study.

We begin with the unstable region, Fig. \ref{pdf1}.

Here, both equations are focusing. It is clear that as the angle increases
there is a significant shift in the mean of the PDF; the stronger the
coupling the more events are produced in both number and magnitude. Thus we
see that oblique interactions can lead to serious rogue events. In fact, for
the focusing scalar case (which is obtained at  $\theta=0$ with $B=0$) the
number of rogue events (events of magnitude greater than three times the
magnitude of the initial condition) is nearly 3\% (2728/100000 events).
Whereas for the coupled case of  $\theta = 10^\circ$  this percentage is
above 22\% (22317/100000 events).

\begin{figure}[tbp]
    \centering
    \includegraphics[width=2.5in]{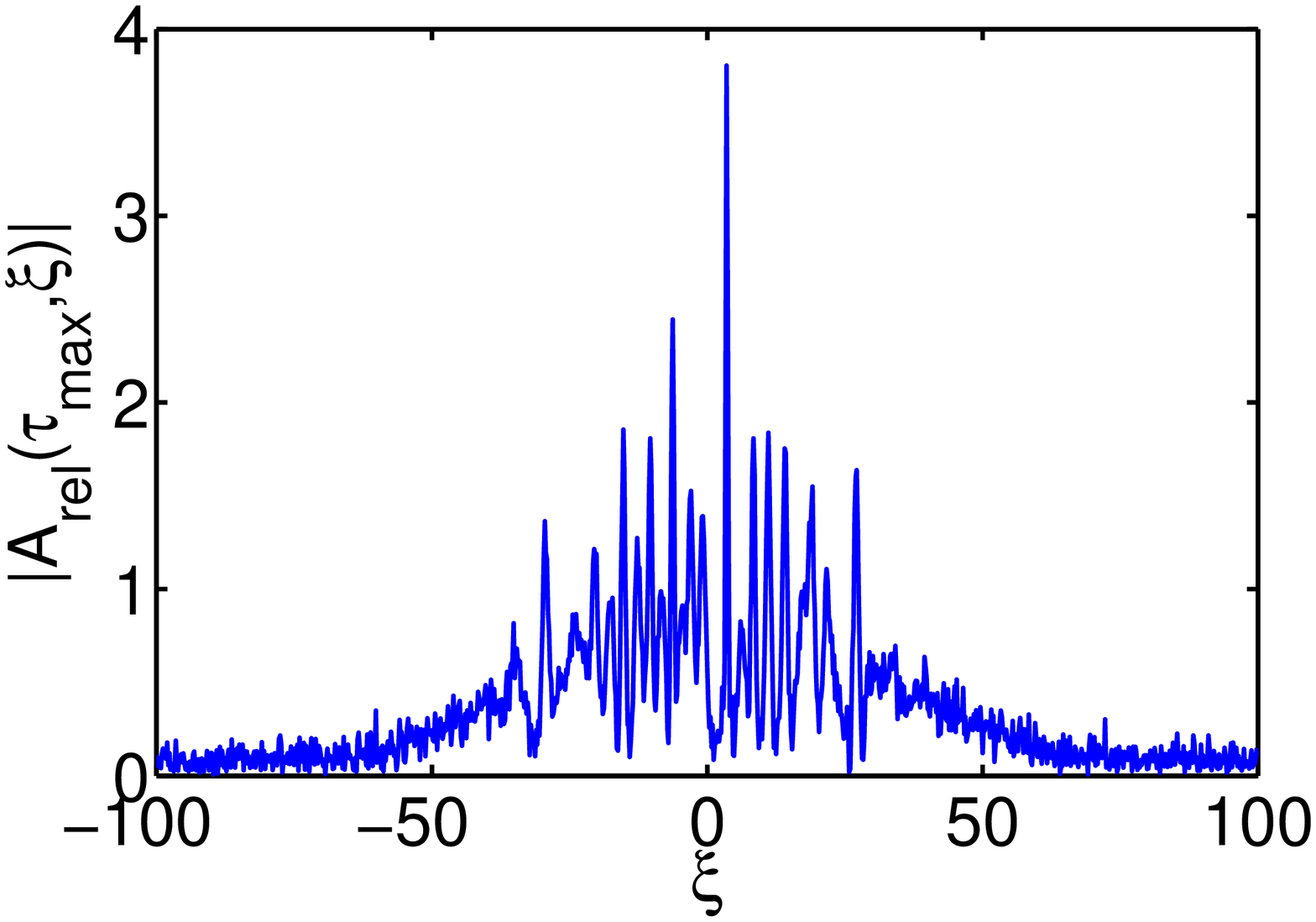}
    \includegraphics[width=2.5in]{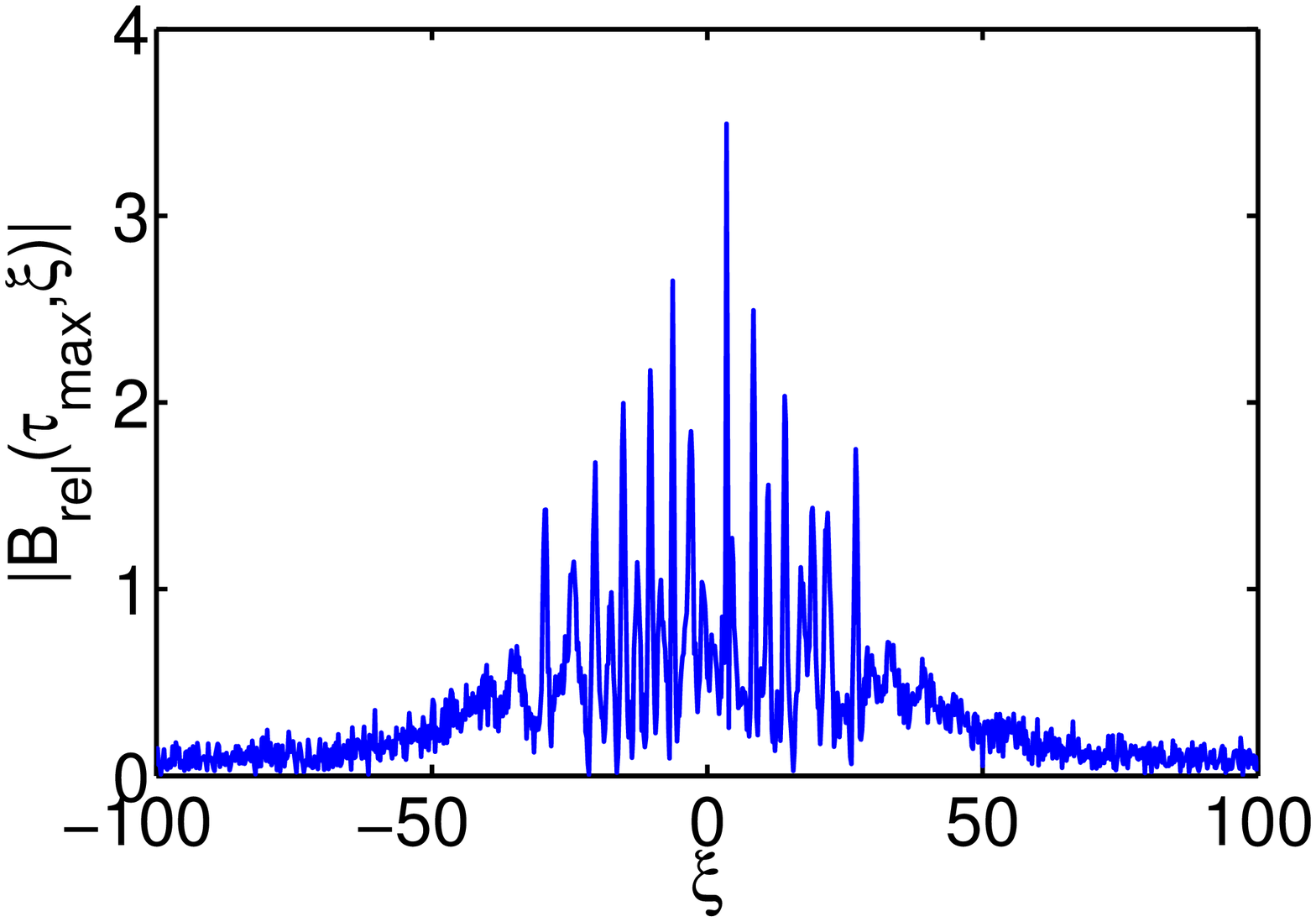}
    \includegraphics[width=2.5in]{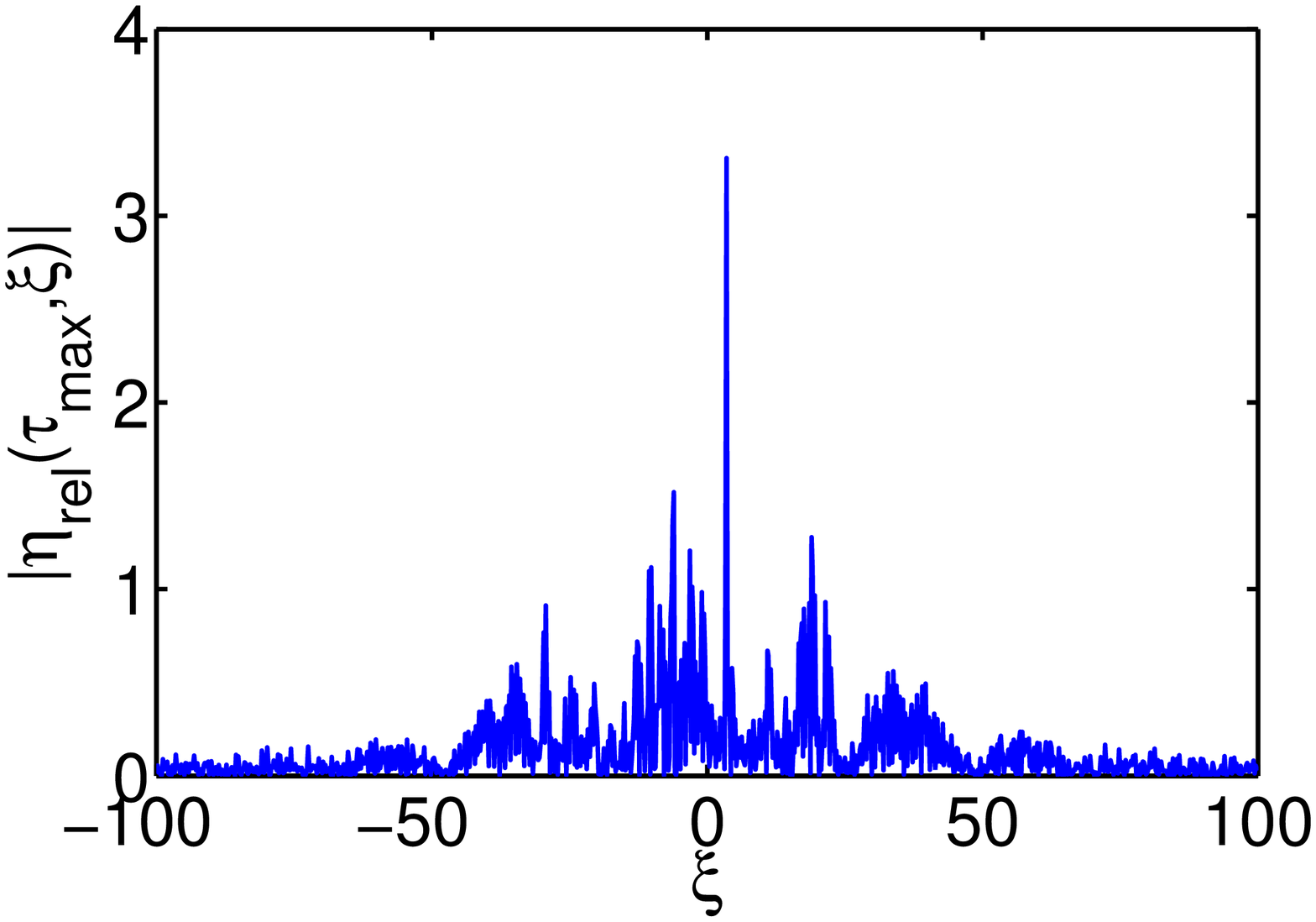}
    \caption{(Color online) A typical rogue event obtained from Eqs.
    \eqref{cnls.final}.
    The
    time $\tau_{\mathrm{max}}$ indicates when the maximum (in amplitude) of this event
    occurs.}
    \label{magnitudes}
\end{figure}

We also demonstrate that both components
play a role for large events; in Fig. \ref{magnitudes} a typical rogue event
is obtained from Eqs. \eqref{cnls.final} and the relative components that
form it. Namely, in the top two figs we obtain $A$, $B$ and plot $A_{\mathrm{rel}}$,
$B_{\mathrm{rel}}$ (i.e. $A$, $B$ divided by their respective initial
conditions).  Similarly from $A$, $B$ we obtain $M$, $N$ and find the wave
elevation $\eta$ and plot $\eta_{\mathrm{rel}}$.

\begin{figure}[tbp]
    \centering
    \includegraphics[width=2.5in]{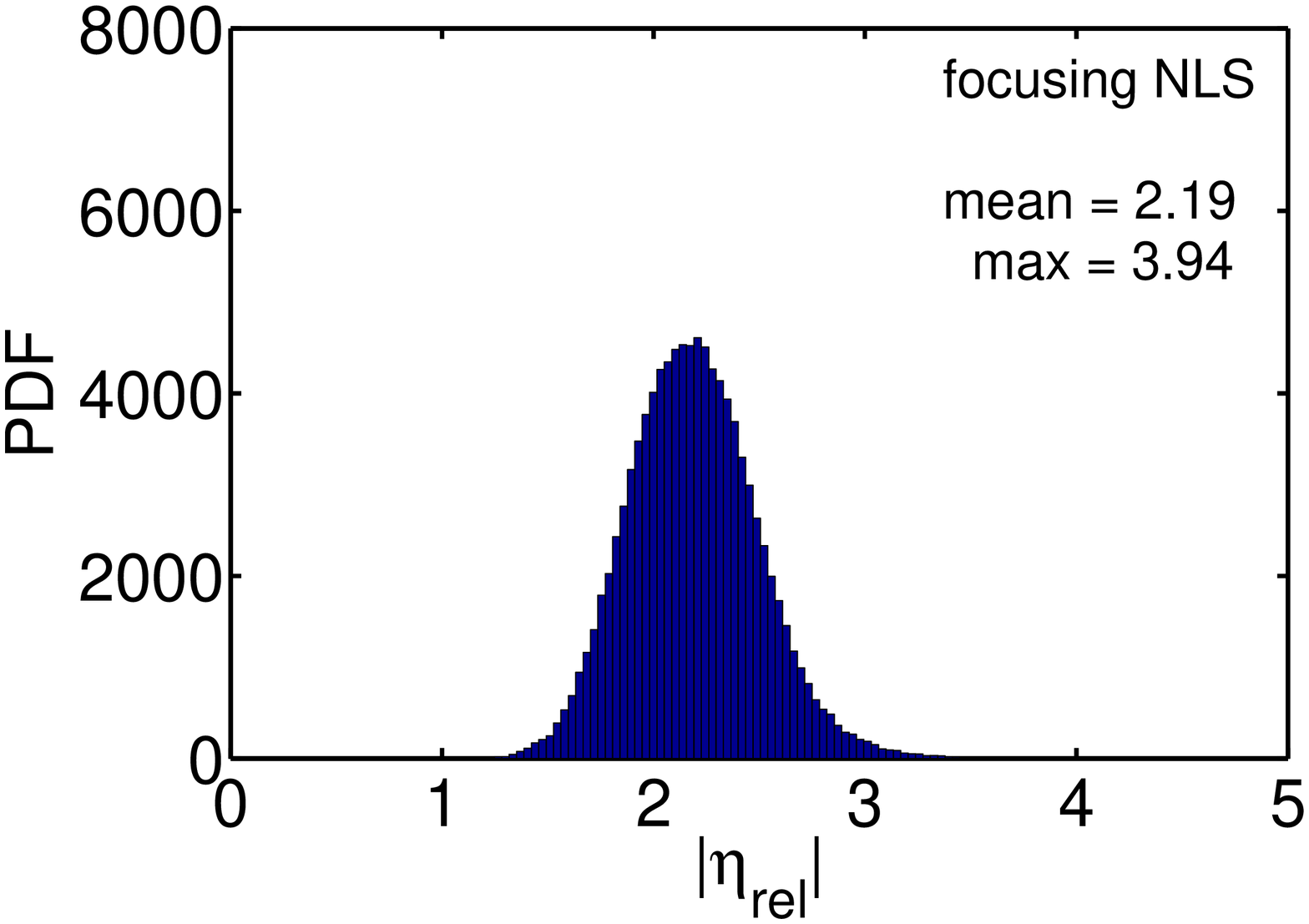}
    \includegraphics[width=2.5in]{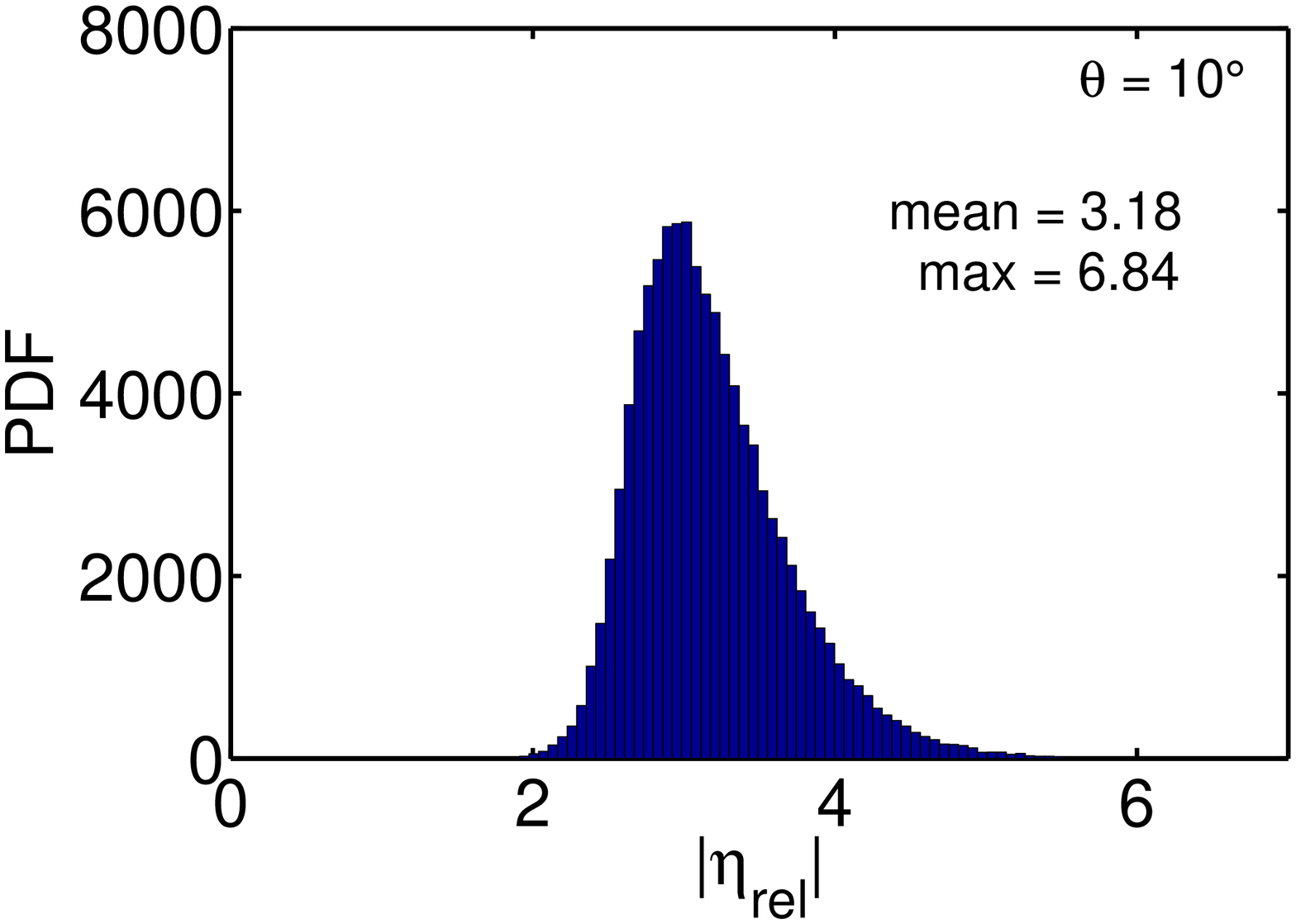}
        \caption{(Color online) The PDFs for the scalar focusing NLS (top) and unstable
        CNLS using JONSWAP data as initial conditions.}
    \label{pdf_jonswap}
\end{figure}

For completeness typical JONSWAP type initial data are also considered
\cite{onorato_report}. No qualitative differences are found as shown in Fig.
\ref{pdf_jonswap}; namely the coupled system leads to far more rogue events
than scalar NLS  equation with these initial conditions. We use the values of the
JONSWAP data from Ref. \cite{onorato_report} along with Eq. \eqref{eta} with $M=N$ at $t=0$ to get the initial conditions for Eqs. \eqref{cnls.final}.

In the scalar NLS equation various modal shapes have been proposed as modes
which describe rogue events, e.g. the Peregrine soliton
\cite{peregrine,akhmediev,chabchoub}. The CNLS system
described here, unlike cases which have different coefficients such as those
considered  in other studies \cite{baronio2}, is not known to be integrable.

\begin{figure}[tbp]
    \centering
    \includegraphics[width=2.5in]{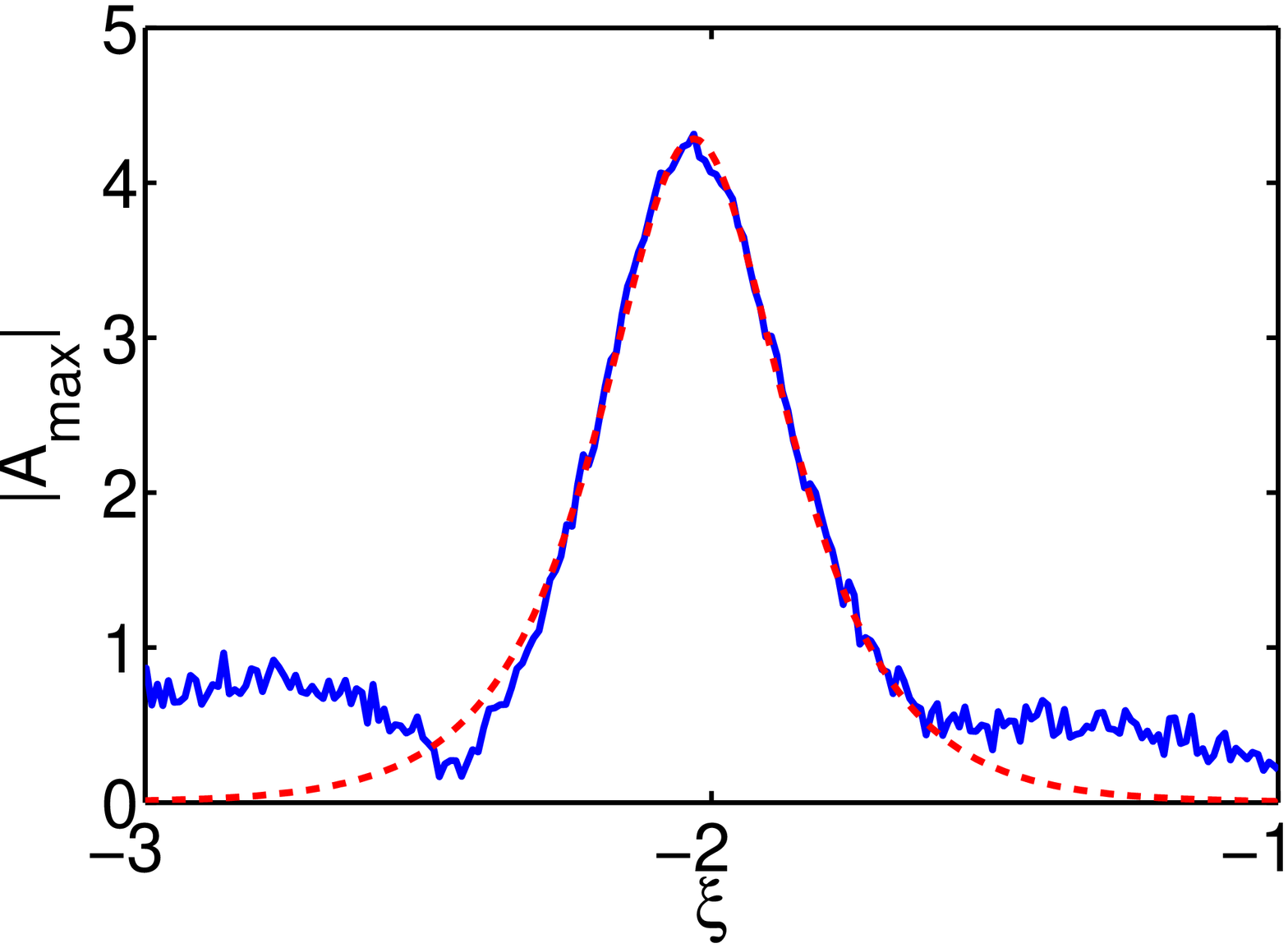}
    \includegraphics[width=2.5in]{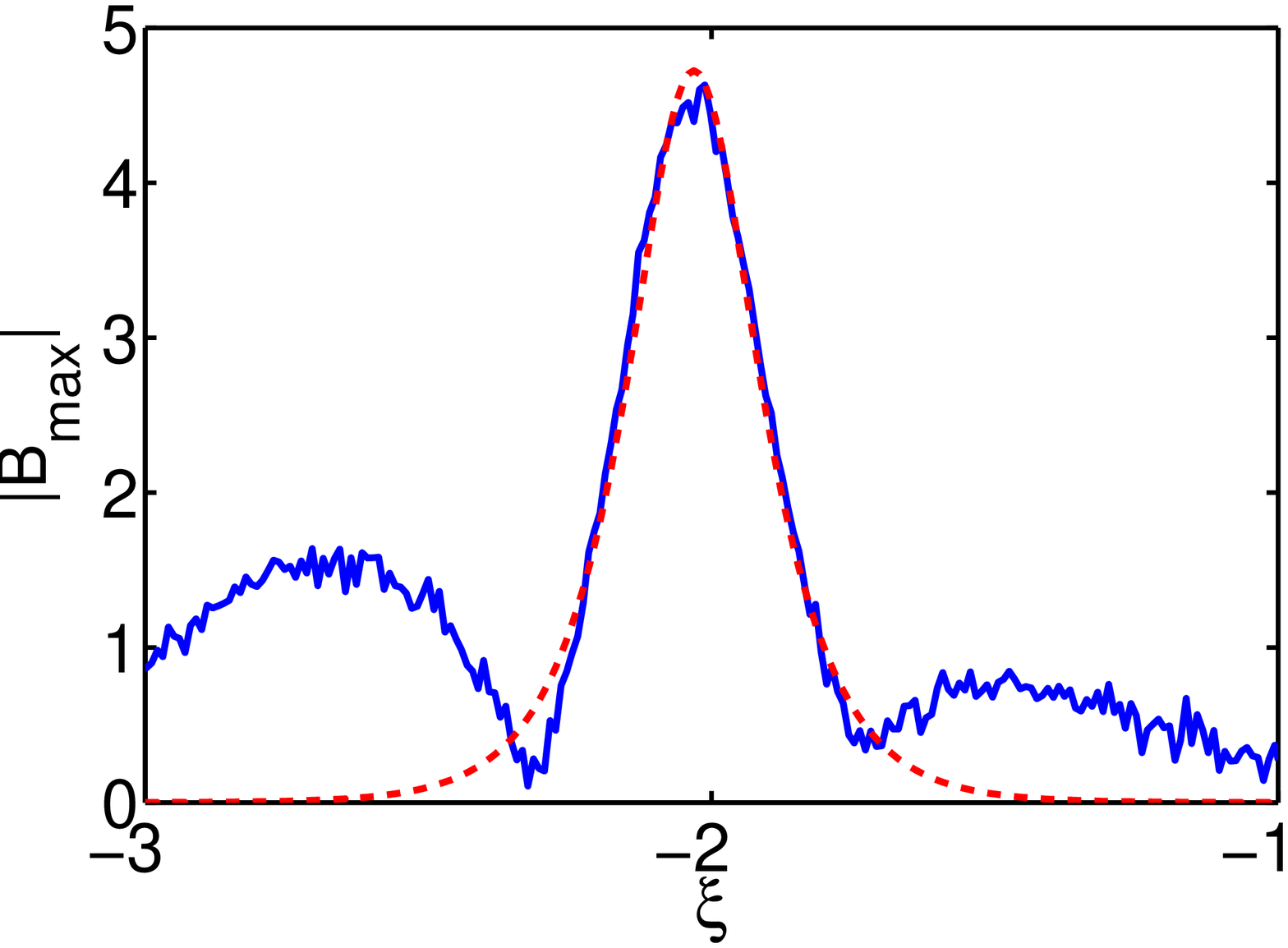}
        \caption{(Color online) A typical zoom-in on a rogue event and the relative
        fit from a hyperbolic secant.}
    \label{rogue_zoom}
\end{figure}

Below, we briefly examine the nature of the extreme rogue waves.
While in the study of rogue phenomena the focus is the surface elevation
$\eta$, the fundamental nature of the solutions of the system \eqref{cnls.final} shifts  attention to the envelopes $A$ and $B$. Consider a typical extreme rogue event much like the one from Fig. \ref{magnitudes}. In order to investigate its properties we zoom in around the maximum height, see Fig. \ref{rogue_zoom}.

\begin{figure}[tbp]
    \centering
    \includegraphics[width=2.5in]{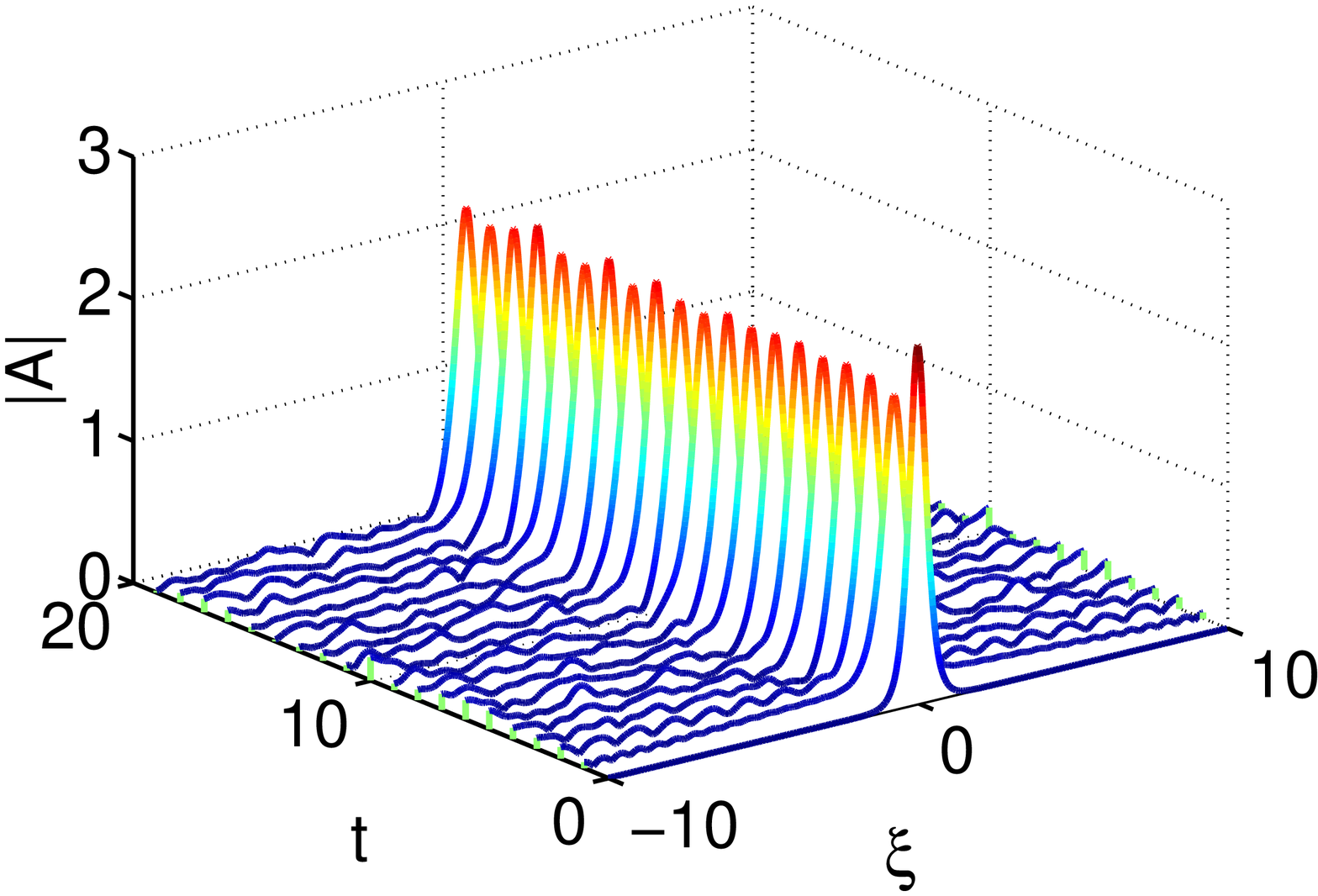}
    \includegraphics[width=2.5in]{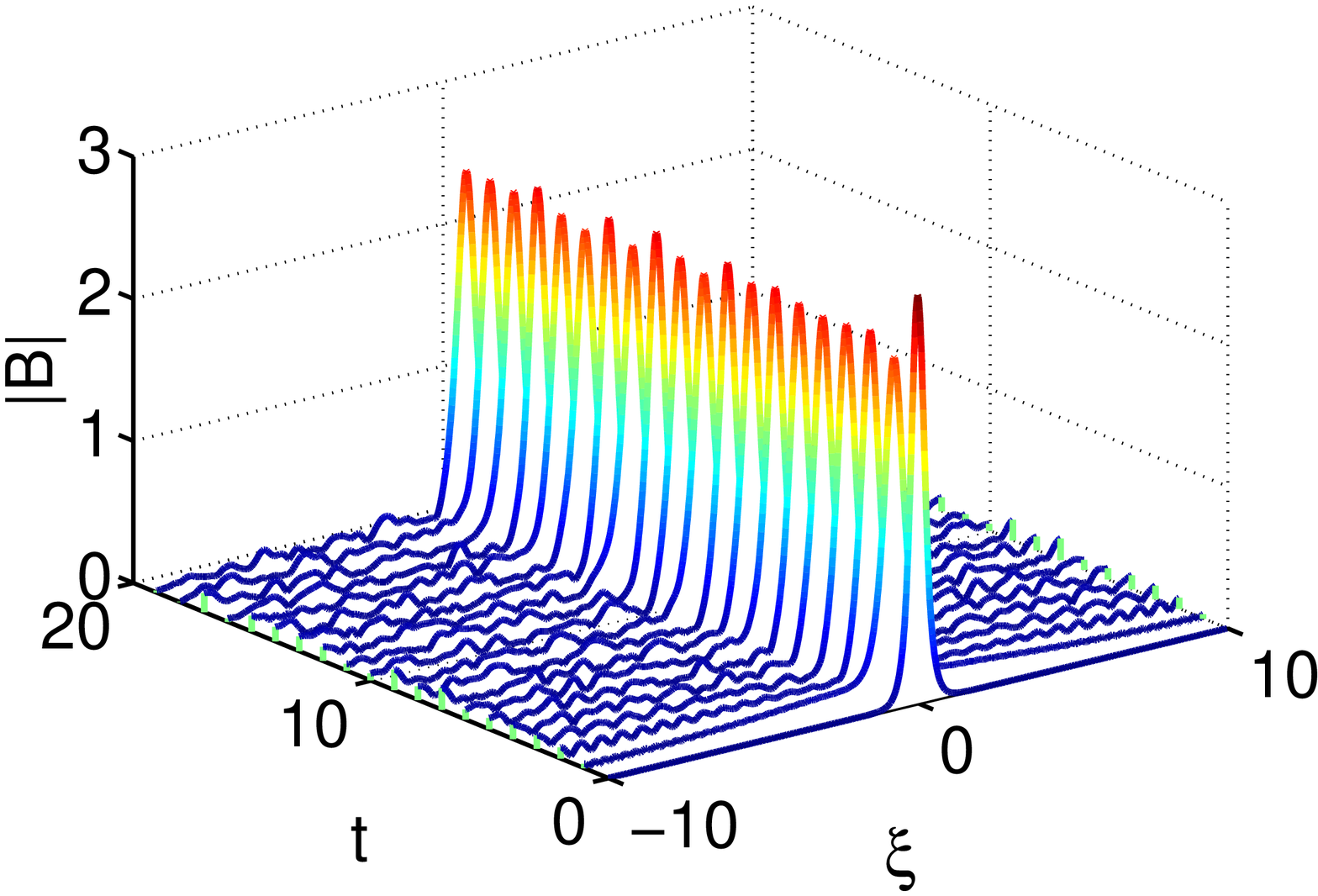}
        \caption{(Color online) The evolution of the envelope components of a
        rogue wave.}
    \label{soliton}
\end{figure}

Then the amplitude of the $A,B$ rogue wave is fitted with hyperbolic secant secant  functions;  we found that this
anzatz gives a very good fit. It should
be noted that these two profiles are not equal (as also seen from the figure) and are
approximated as
\[
\
A(\xi)=4.284\; \mathrm{sech}(6.852\xi),\;\;
B(\xi)=\; 4.724\;\mathrm{sech}(9.728\xi)
\]
Next we evolve we use these profiles as, initial conditions for the CNLS system, Eqs. \eqref{cnls.final}; we
depict the evolution in Fig. \ref{soliton}.

Remarkably, the initial condition holds its shape and position and propagates as
an approximate solitary wave --soliton solution of the system. The small amount of radiation shed is to be expected as this is only
an approximate solution to the system. However, it also attests to its stability as it
does not break down or disperse as one may expect from a function which is not near a true solution. This sheds new light into the
properties of rogue waves as they evidently can be  solutions of  what is not known/expected to be an integrable
system. Similar results are obtained for other rogue events. Further comparisons and more detailed analysis will be carried out  in a future communication.

\begin{figure}[tbp]
    \centering
    \includegraphics[width=2.5in]{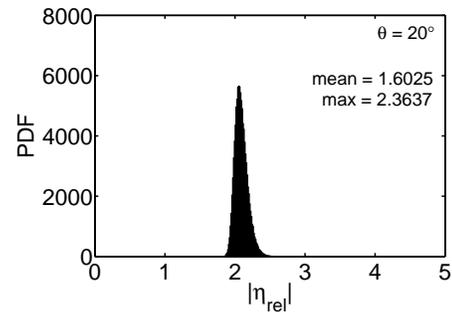}
    \caption{(Color online) The PDFs for the weakly unstable CNLS.}
    \label{pdf2}
\end{figure}

We continue with the weak unstable region, Fig. \ref{pdf2}. In this region
one of the equations is focusing while the other is defocusing. Note here
that for the CNLS system one still has unstable MI growth from the initial
condition. However, in this case it produces fewer rogue events than that of
the single scalar NLS equation--compare with Fig. \ref{pdf1} (top). As
mentioned in the introduction it is important to note that despite the fact
for this choice of parameters, the growth rate is greater in the CNLS system
than that of the corresponding scalar NLS equation \cite{kourakis} there are
fewer not more rogue events generated in  the coupled NLS system.

From the above analysis it is clear that the CNLS system not only corresponds
to an interesting and physically relevant situation in deep water but it also
produces novel and significantly different rogue phenomena than its scalar
NLS counterpart. It is important to note that in certain parameter regimes a
major increase in the number of rogue events can be produced from the CNLS
system as compared with the scalar NLS equation. While rogue events are
closely associated with MI, nevertheless larger MI growth rates do not always
lead to more rogue events. Furthermore, these events are approximate solitary wave--soliton
solutions of the coupled NLS system which is not known to be integrable. It appears unlikely that these
solitary--soliton waves can be rational solutions since the latter are typically limits of multi-soliton solutions which are
intimately related to integrable equations \cite{ist}.

To summarize, an interacting coupled NLS wave envelope system associated with
the Euler equations in deep water with different group velocities is studied.
In general, this system of equations exhibits modest nonlinear interaction
since the group velocity terms would lead to the separation of localized
states. However this coupled system can be reduced to the CNLS system by
adding an additional restriction. This links the angle of interaction with
the group velocities of the interacting waves. With this additional
condition, referred to here as the rogue condition, it is found that
depending on the angle of interaction, the coupled system can substantially
enhance the number and size of rogue events as compared to the scalar NLS
equation which itself has been linked to rogue wave phenomena. The type of
rogue wave obtained from the CNLS system is vectorial in nature and as such
is fundamentally different from the rogue events in the scalar NLS equation.

\begin{acknowledgments}
We thank Anastasios Raptis for his help with parallel programming. This
research was partially supported by NSF under grants DMS-0905779 and
DMS-1310200.
\end{acknowledgments}

\end{document}